# Emergent properties in supercrystals of atomically precise nanocluster and colloidal nanocrystals

Korath Shivan Sugi*,[a], Andre Maier*,[a] and Marcus Scheele [a]

We provide a comprehensive account of the optical, electrical and mechanical properties that emerge from the self-assembly of colloidal nanocrystals or atomically precise nanoclusters into crystalline arrays with long-range order. We compare the correlation between the supercrystalline structure and these emergent properties with similar correlations in crystals of atoms to address the hypothesis that nanocrystals and nanoclusters exhibit quasi-atomic behaviour. We come to the conclusion that, effectively, this analogy is indeed justified, although the chemical origin for the same emergent properties are substantially different in crystals of atoms vs. supercrystals. We provide an outlook onto the most promising applications of supercrystals of nanocrystals and nanoclusters and discuss the challenges to be overcome before their commercialization.

## 1 Outline and scope

"Analogy Between Atoms in a Nanocrystal and Nanocrystals in a Supracrystal: Is It Real or Just a Highly Probable Speculation?" is the title of a perspective article by Goubet and Pileni from 2011.[1] In the present feature article, we respond to this question by discussing new results that have been reported since then, in particular the appearance of emergent optical, electrical or mechanical properties in such "supra- or supercrystals" (SCs). The building blocks of these SCs, for which the "quasi-atomic" analogy has been postulated, are colloidal nanoparticles which are capped with ligand molecules and can thus be dispersed in solution (**Figure 1**). Their strongly size-dependent optoelectronic properties in combination with a solution-processed fabrication offer a powerful platform for the development of tunable optoelectronic devices [2-4] such as transistors,[5, 6] solar cells,[7, 8] light emitting devices,[9, 10] photodetectors[11, 12] and thermoelectric power generators.[13] First examples of SCs were demonstrated more than 25 years ago[14] and since then, a large number of excellent reviews has covered the structural aspects of this material class and the chemical tools to tailor them. Here, we focus on the fundamental question to which extent the chemical and physical concepts developed for atomic crystals also apply to SCs, that is, to which degree nanoparticles can really be seen as "quasi-atoms". When atoms are self-assembled into a periodic array, this atomic crystal exhibits properties which are much more than the mere sum of its building blocks. To name just a few examples, diamond exhibits greatly different electrical, optical, and mechanical properties than graphite owing solely to the different crystalline structure of the same constituting building blocks. The physical chemistry at the surface of an atomic crystal is vastly distinct from that of its bulk interior due to a lower coordination of surface atoms. Crystalline silicon with long-range structural order is a greatly different material than amorphous silicon without any structural periodicities. Electric transport in black phosphorus is highly anisotropic due to the different structural motifs in the so-called armchair vs. zigzag direction.[15-17] We elaborate here that similar observations of "emergent properties" have recently also been made for SCs. These findings are not only strongly affirmative of Goubet's and Pileni's question from 2011, but they also highlight the exciting perspectives for chemists to exploit structure-property correlations in SCs.

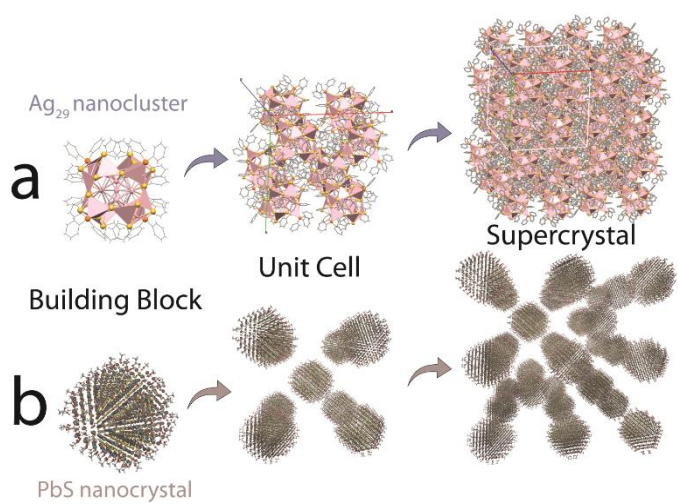

**Figure 1.** Self-assembly of **a)** nanoclusters, exemplified here by Ag$_{29}$(1,3-benzenedithiol)$_{12}$(triphenylphosphine)$_4$ (Redrawn from ref. 98 and **b)** nanocrystals, exemplified here by acetic acid-capped PbS, into supercrystals with periodically repeating unit cells.

This article is structured as follows: we begin with an introduction of some of the fundamentals of SCs, specifically a definition and description of the building blocks (2.1), a short overview of the chemical tools for their self-assembly into SCs (2.2), followed by the description of charge transport in SCs and chemical tools to tailor it (2.3). In section 3, we present evidence of emergent properties of SCs that have so far been reported, and we deliberately limit ourselves to optical (3.1) as well as electrical properties (3.2). Section 4 contains examples of our own recent work on structure-property correlations in SCs that we view as analogues to the selection of emergent properties in atomic crystals stated above. Section 5 reviews current challenges, possible solutions and opportunities of this field, followed by a short conclusion in section 6.

## 2 Fundamentals

### 2.1 Building blocks

Nanoparticles are nanometer-sized fragments of metals, semiconductors, dielectric or organic materials and are extensively researched for applications in diverse fields such as biology, medicine, and optoelectronics.[18] Nanoparticles have significantly different properties than the respective bulk materials. The surface area-to-volume ratio becomes larger the smaller the solid under consideration, which leads to significantly increased surface energies and reactivities.[19] The tremendous developments of the past decades have enabled finely tuned syntheses for the realization of nanoparticles of various materials, sizes, and shapes.[18, 20] One growing field of research is the use of nanoparticles as versatile building blocks for the self-assembly into regular crystal-like materials, often referred to as "supercrystals (SCs)",[14, 21] with properties by design.[22-24] Once assembled, SCs are expected to exhibit emergent collective properties, caused by quantum mechanical and dipolar coupling between these individual building blocks.[2, 19, 25-28] In this article, we will focus on two types of building blocks, namely atomically precise nanoclusters (NCLs) with diameters on the order of 1–3 nm as well as semiconductor nanocrystals (NCRs) with diameters of up to 20 nm. One prerequisite for the self-assembly into ordered SCs is a sufficiently narrow size distribution of <10% for the NCRs, whereas in the case of atomically precise NCLs (e.g. no size distribution), a high molecular purity is required.[14, 29-32]

### 2.1.1 Atomically precise nanoclusters

In contrast to the inherent polydispersity of NCRs, NCLs, which are often considered a bridge between molecules and macroscopic materials,[32, 33] are atomically precise, rendering them ideal building blocks for SCs with long-range structural order. NCLs are protected with ligands such as phosphines, selenols, thiols, alkynyl, etc. Their composition can be precisely written as $M_xN_y$ (x-number of metal atoms, y-number of protecting ligands).[34] Recent advances in synthesis strategies and structure determination enabled a better understanding of structure-property correlations.[35, 36] The structure of thiolate-protected NCLs consists of an inner metal core, held together by metal-metal bonds, which are stabilized by staple motifs, and an outer ligand shell as suggested by the "divide and protect model".[33, 37] The size of an NCL is comparable to its exciton Bohr radius, enabling significant quantum confinement, which results in a discrete electronic structure. This leads to unique optoelectronic properties such as multiple absorption bands, nonlinear properties, photoluminescence, hypergolic properties etc.[32, 38-40] The properties of NCLs can be tuned by altering individual core atoms and their packing. As such, NCLs find application in catalysis, sensing and optoelectronic devices.[32, 38, 41-43]

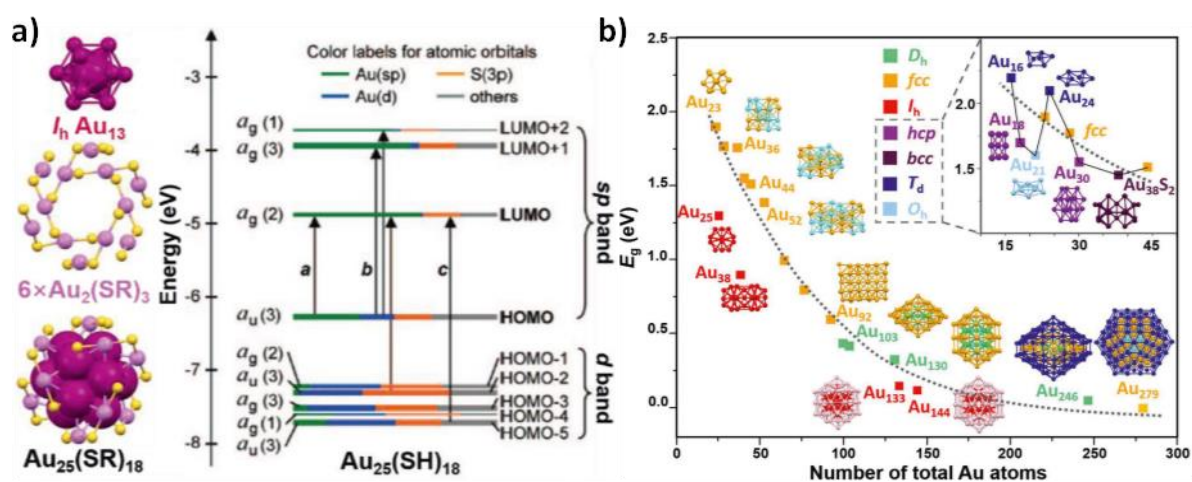

**Figure 2. a)** Structure anatomy and Kohn–Sham (KS) orbital energy level diagram of $[Au_{25}(SC_2H_4Ph)_{18}]$ NCL, colour label: magenta = Au, yellow = S. **b)** The bandgap of gold NCLs as a function of the total number of metal atoms. Inset: Bandgap of scalable NCLs with atomic packing of the kernel. Figure adapted with permission from ref. 47.

Owing to their small size of less than 3 nm, NCLs behave like molecules with electron transitions/excitonic states, unlike plasmonic nanoparticles, which exhibit surface plasmon resonances.[44] In 2008, Jin and co-workers reported the first study on the electronic properties of thiol-stabilized $Au_{25}$ NCLs, which were found to be strongly quantum confined.[45] These $Au_{25}(SR)_{18}$ NCLs consist of an $Au_{13}$ icosahedral core protected by six dimeric staple motifs (**Figure 2a**). Time-dependent density functional theory (TDDFT) calculations implied that the molecular-like behaviour of $Au_{25}$ NCLs arises from both core and motif-based orbital transitions. The highest occupied molecular orbital (HOMO) and the lowest unoccupied molecular orbital (LUMO) have contributions from $Au_{13}$ core orbitals, whereas HOMO-n orbitals have contributions from core and staple motifs (**Figure 2a**). Due to the presence of metal-containing staple motifs, thiolate-protected NCLs behave differently than phosphine-protected NCLs, which do not exhibit such motifs. Recently, Jin and co-workers systematically probed the evolution of thiolate-protected gold NCLs from the molecular to the metallic/plasmonic state.[46] They compared the optical properties of a series of NCLs with their size/structure and categorized them into three groups; non-scalable NCLs, medium-sized NCLs, and scalable NCLs (**Figure 2b**). Non-scalable NCLs consist of less than 50 core atoms, and their bandgap as well as the optical properties are determined by structure rather than their size. In the case of the second group, the bandgap and carrier dynamics significantly depend on size, whereas the absorption spectra depend on their atomic structure. The optical properties of scalable NCLs having more than 100 core atoms rely mainly on their size. As the size increases further, they begin to show plasmonic optical properties. The evolution of the optical properties of thiolate-protected silver and alloy NCLs is even more complex.[47] $Ag_{141}$, $Ag_{206}$, and $Ag_{347}$ NCLs are plasmonic with a distinct SPR band, whereas $Ag_{146}$ NCL was confirmed to be non-metallic.[48-50] Alloying silver into Au NCLs increases the energy gap.[51] Details have been summarized in several reviews.[51-53]

### 2.1.2 (Monodisperse) nanocrystals

Colloidal nanocrystals (NCRs) are crystalline fragments of inorganic materials with diameters of ~ 2–20 nm and consist of approx. ~$10^2$–$10^4$ atoms.[20, 29] Typically, long-chain hydrocarbons are used as ligand molecules whose functional head groups bind to the surface of the NCR.[54] This passivation prevents agglomeration of the NCRs due to steric stabilization and allows stable dispersion in nonpolar solvent.[55] In contrast to NCLs, NCRs exhibit an inherent size distribution, meaning that they are never exactly identical in size, i.e. number of atoms, and structure. The most homogeneous NCRs today still feature a size distribution of ~4% (**Figure 3a**).[22, 56-58] The properties of the NCRs can easily be tuned by slightly adjusting the synthesis parameters, such as temperature or time, to yield the NCRs of the desired size[22, 56] or composition.[59] Further, the shape and faceting of the NCRs can be well controlled to specifically fabricate the desired building block with their unique optoelectronic properties.[60-64]

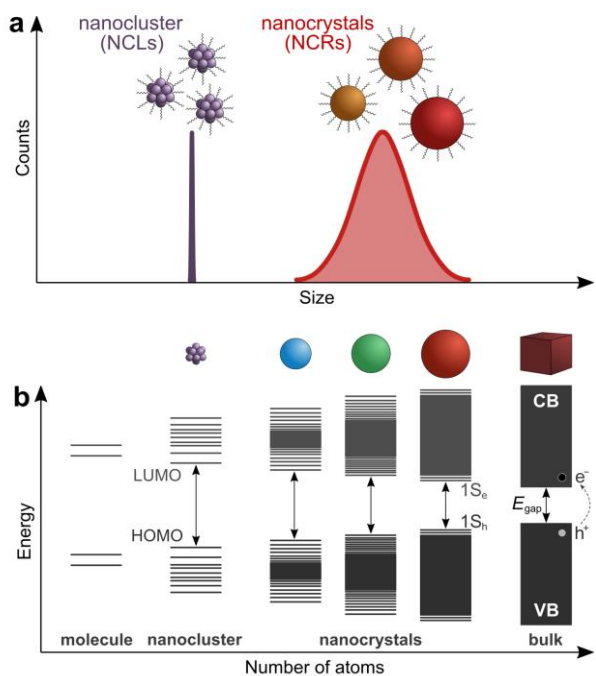

**Figure 3. a)** Schematic illustration of the size distribution of nanoclusters (NCLs), featuring a sharp delta function, and nanocrystals (NCRs), showing a Gaussian size distribution. **b)** Illustration of the quantum confinement effect of nanoclusters and semiconducting nanocrystals. Figure adapted with permission from ref. 255.

The tremendous developments of the past decades[18, 20] have enabled finely tuned syntheses for the realization of NCRs of various materials, sizes and shapes – ranging from semiconducting PbS/PbSe[56, 65-67] and CdSe NCRs[14, 57, 68] over easily tunable lead halide perovskite NCRs[59, 69-72] to silicon,[73] $Fe_3O_4$[74] and Au or Ag NCRs[75] or even quasi-2D nanoplatelets[76, 77] – to name just a few examples.

**Semiconductor NCRs & quantum confinement:** Semiconductor NCRs are a subset of NCRs and commonly consist of II-VI, III-V, and IV-VI semiconductors, prepared by means of inexpensive and scalable solution-processed synthesis.[4, 22, 29, 69] Semiconductor NCRs are often referred to as quantum dots, indicating the occurrence of size-dependent quantum effects, which dramatically alter their optical and electronic properties.[2, 78] In NCRs with sizes smaller than their exciton Bohr radius, the excitons are spatially constricted in three dimensions, which is called quantum confinement.[29] Consequently, the band gap of NCRs increase with decreasing particle size due to strong confinement, and the electronic states become more discrete (**Figure 3b**).[26, 79] The emerged CB and VB band edges are termed $1S_e$ and $1S_h$ for electrons and holes, respectively.[29] The effect of quantum confinement gives rise to unique optical and electronic properties of colloidal semiconductor NCRs, as the energy levels and both, the absorption and emission wavelength, can be tuned by changing the size of the NCRs. However, due to size-dependent quantum confinement and their inherent size distribution, NCR ensembles exhibit a distribution of energy levels, which is a distinct challenge for applications where fully homogeneous electronic properties are mandatory.

**Metallic NCRs & particle plasmons:** Another subset of NCRs are metallic NCRs.[60] Bulk metals and large metal NCRs have a continuous band of energy levels, which results in plasmonic states.[80, 81] Incoming light can induce an electrical dipole by creating a collective oscillation of the conduction electrons with respect to the atomic cores.[82, 83] Due to this localized surface plasmon resonance (LSPR), a strong electric field enhancement can be created on the surface of metallic NCRs. The corresponding LSPR frequency depends on the NCR composition, size and shape, as well as the surrounding medium.[82]

### 2.2 Self-assembly into supercrystals

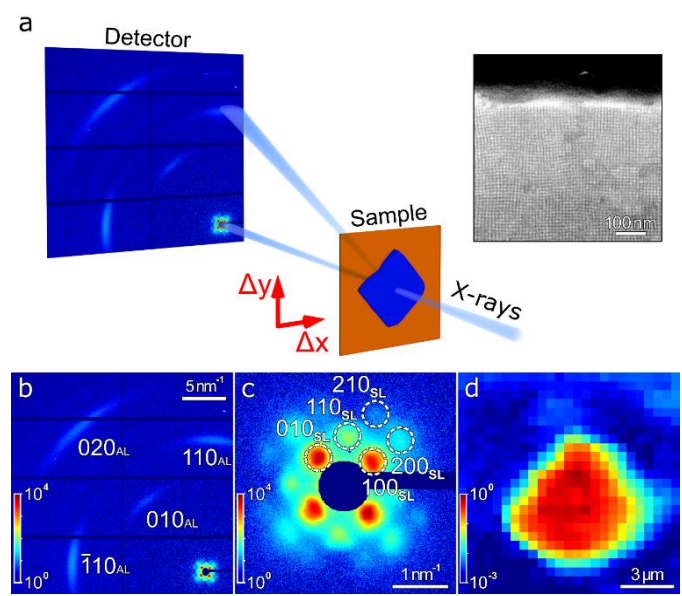

**Figure 4. (a)** X-ray nanodiffraction of a $CsPbBr_2Cl$ supercrystal (see top right for the electron micrograph) in transmission geometry. (**b**) Average wide- and **c)** small-angle diffraction patterns (**d**) X-ray micrograph based on the integrated small-angle scattering intensity with a step size of 500 nm. Figure reproduced with permission from ref. 228.

SCs of NCRs or NCLs result from the slow self-assembly of the building blocks in section 2.1 under thermodynamic conditions, that is, if kinetic arrest is avoided.[84] SCs that additionally exhibit a uniform orientation of their building blocks with respect to the overall SC are termed "mesocrystals" with many naturally occurring examples.[85, 86] Since most of the building blocks in section 2.1 contain heavy elements, X-rays are strongly scattered by SCs, and X-ray scattering methods are excellently suited to study their structure.[87] While thin films of SCs are best examined with grazing incidence techniques due to their high surface sensitivity,[88] three-dimensional SCs are better studied in transmission geometry due to the smaller footprint.[88] Specifically, mesocrystalline assemblies of NCRs, which exhibit an extended (atomic) lattice themselves, are characterized by precise angular correlations between the two lattices, which may be analyzed by X-ray cross-correlation analysis (**Figure 4**).[89] Recent advances at synchrotron radiation sources have enabled "nanodiffraction" experiments, utilizing a beam focus of 30 nm x 30 nm[90] or even smaller.[91] While nanodiffraction has greatly improved the ability to study individual SCs and obtaining single-crystalline diffraction patterns,[92, 93, 94] advances in the crystallization of NCRs or NCLs into SCs have yielded increased single-crystalline domain sizes, which can reach up to 100 µm² for NCRs and up to mm² for NCLs.[95-97]

The main synthetic procedures toward SCs have been reviewed extensively,[24, 36, 55, 78] and they entail solvent evaporation by drop-casting,[98] dip-coating or spin-coating, solvent destabilization, e.g., by vapor diffusion,[99] confinement in an emulsion droplet, liquid-liquid interface preparation, electrocrystallization[97] and decomposition of surfactant micelles (**Figure 5**). According to Boles et al., these procedures can roughly be classified into two groups, namely the nucleation in bulk solution vs. that under 2D confinement.[55] One of the most noteworthy differences between SCs vs. atomic lattices is the presence of a soft molecular shell on the surface of the nanocrystals or -clusters. The flexibility of this ligand shell induces a variety of entropic effects, which are absent in atomic lattices and rather resemble the behaviour of micelles or globular proteins.[100] Ligand-ligand interactions, such as electrostatic interactions, hydrogen bonding, and van der Waals interactions between the peripheral ligands greatly affect NCL and NCR assembly and dictate the final SC structure.[36, 101-103] This is well demonstrated in the case of $Ag_{44}$ or $Au_{12}Ag_{32}$ NCLs.[104] When the protecting ligands are 3,4-difluorothiophenol ($HSPhF_2$) or 4-fluorothiophenol (HSPhF), the NCLs assemble into triclinic SCs,. In contrast, assembly into a monoclinic unit cell takes place when 4-(trifluoromethyl) thiophenol ($HSPhCF_3$) was used. Ligand-ligand interactions are also deemed responsible for orientational order in SCs, that is, the degree to which all building blocks are isooriented with respect to the SC.[105]

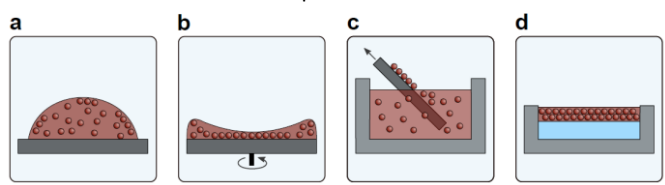

**Figure 5**. Schematic illustration of evaporation-based self-assembly methods: (**a**) drop-casting, (**b**) spin-coating, (**c**) dip-coating, (**d**) self-assembly at the liquid/air interface. Figure reproduced with permission from ref. 255.

Besides these interparticle interactions, intraparticle interactions also need to be considered, which enables the patterned arrangement of ligands around the particle core. The interactions vary depending on the functional group of the protecting ligands. Aromatic ligands mainly utilize π···π and CH···π interactions whereas hydrogen bonding is predominant in carboxylate ligands. An excellent example for ligand-directed assembly and the role of intra- and interparticle interactions is the case of $Au_{103}$,[106] $Au_{246}$,[107] and $Ag_{44}$ NCLs.[108] In the case of naphthalenethiol-protected $Au_{103}$ NCLs, the two-naphthalene ligand forms a T-shaped dimer via intracluster C−H···π interactions. The two pairs of such T-shaped dimers form a tetramer via intercluster C−H···π interactions, which leads to a herringbone-like assembly of ligands and directs NCLs to assemble in a monoclinic lattice. In the case of $Au_{246}$ NCLs, the MBT (*p*-methylbenzenethiolate) ligands self-organize in two different patterns at the pole and waist of the NCL surface via intracluster C−H···π interactions. Further, intercluster interactions result in their hierarchical assembly. The $Ag_{44}$ NCLs were protected with p-mercaptobenzoic acid (p-MBA). The intracluster and intercluster interactions involve hydrogen bonding between doubly bundled and triply bundled ligands, resulting in a triclinic unit cell.

Another important difference between atoms and NCRs or NCLs is that the latter are often not spherical and can exhibit complex polyhedral shapes. The self-assembly of such polyhedral objects has been described as the result of a many-body directional entropic force towards a preferential alignment of flat facets.[109] Even more complex SC structures are possible upon self-assembly of at least two different building blocks to yield binary or multinary assemblies.[31, 110]

In general, assemblies of NCRs exhibit higher symmetries and less polymorph diversity than SCs built from NCLs. For the former, the predominant SC structures are face- or body-centered cubic as well as hexagonally close-packed,[105] while the latter are typically monoclinic or triclinic, together with some rare examples of cubic and trigonal lattices. An example for the rich polymorphism of NCLs are $Ag_{29}$ NCLs, which were shown to result in cubic[98] or trigonal[99] superlattices by varying the crystallization protocol. Similarly, changing the solvent from dimethylformamide (DMF) to N-methyl-2-pyrrolidone (NMP) and tetramethylene sulfone (TMS) resulted in orthorhombic and cubic crystals.

The symmetry of the NCL kernel (i.e., the arrangements of metal atoms) has also been shown to dictate the NCL assembly in that a higher kernel symmetry usually invokes a higher symmetry of the SC.[111] The most prevailing atomic packing observed in NCL assemblies are ABAB and ABCD packing modes.[32, 112] Recently, a rare ABCDEF-6H packing mode was reported for $Ag_{60}$ NCLs.[113]

Counterions can dictate the assembly of charged NCLs. For instance, $[Au_{21}(SR)_{12}(PCP)_2]^+$ (R = cyclohexyl; PCP = bis(diphenylphosphinomethane)) NCLs with a $AgCl_2^-$ counterion crystallize into a triclinic lattice whereas with $Cl^-$ as the counterion, a monoclinic lattice was obtained.[114] It is believed that the phenyl rings in this NCL exhibit a strong interaction with the counterion, leading to a distorted T shape configuration for $AgCl_2^-$ ions, whereas those with $Cl^-$ ion result in a parallel displaced configuration, thereby directing their packing into different lattices. Recently, S. Chen et al. assembled $[Au_1Ag_{22}(S-Adm)_{12}]^{3+}$ NCLs into cubic and hexagonal lattices by controlled addition of $SbF_6^-$ counterions.[115] L. He et al. demonstrated the layer-by-layer assembly of $[Ag_{26} Au(2-EBT)_{18} (PPh_3)_6]^+$ and $[Ag_{24} Au(2-EBT)_{18}]^-$ NCLs into a triclinic lattice. Here, NCLs of opposite charge can be viewed as counterions that guide their assembly.[116]

Similar to atomic lattices, a core research interest has been the understanding of defects and their formation in SCs.[95, 117] A key factor in their formation is the contraction of the SC during evaporation of the solvent.[93, 118] This contraction induces isotropic strain, favouring the occurrence of point defects at first, from which more complex defects (dislocations, twin boundaries, precipitates, etc.) can follow.[118] Specifically for NCRs, this defect formation is additionally facilitated by a size mismatch of adjacent NCRs. After assembly into an SC, further structural transformations, such as necking, oriented attachment and epitaxial connection between adjacent NCRs or NCLs are possible.[119] While such oriented attachment on the large scale can lead to new materials with potentially very exciting properties,[114, 120, 121] managing defects in these materials is key toward their full exploitation.[122] In this regard, the formation of dimers has been identified as a central first step,[123] and it has been argued that classical bulk crystal dislocation theory is successful in predicting the result of oriented attachment in an SC with certain defects.[124] This is another demonstration of the many analogies between atomic lattices and SCs.

## Co-crystallization of mixed building blocks into supercrystals

Co-crystallization is a widely applied method in the pharmaceutical sciences.[125] Recently, the co-crystallization of atomically precise NCLs into the same SC has received significant attention,[34, 126] and supramolecular interactions have been identified as the main driving force for this action. A few examples for co-crystals of atomically precise thiolate-protected NCLs are summarized in **Table 1**. Typically, co-crystallization occurs between building blocks of similar size or structure, which is evidenced by the examples in row 1-5. Specifically, in the $Ag_{40}$/$Ag_{46}$ NCL co-crystal (row 4), the outer shell ($Ag_{32}(DMBT)_{24}(TPP)_8$) is identical but the inner metal kernels ($Ag_8$ and $Ag_{14}$) are different.[127] In row 1, the two building blocks differ only in the occupancy of the center position, such that their outer shells are indistinguishable. Conversely, in the examples in row 3 and 5, the metallic kernel is the same, but the outer shells differ.[128, 129] In contrast, Yan et al. reported the unusual co-crystallization of $(AuAg)_{45}(SphMe_2)_{27}(PPh_3)_6$ and $(AuAg)_{267}(SphMe_2)_{80}$ NCLs, which exhibit large size differences (row 6).[130] Each $(AuAg)_{267}$ NCL is coordinated by six neighboring $(AuAg)_{45}$ NCLs, and it is believed that intercluster interactions play a major role in this assembly.

Apart from supramolecular interactions, electrostatic interactions also play a significant role in the co-crystallization of NCLs. He et al. reported a double nanocluster ion compound where the cationic $[Ag_{26}Au(2\text{-}EBT)_{18}(PPh_3)_6]^+$ and anionic $[Ag_{24}Au(2\text{-}EBT)_{18}]^-$ NCLs assemble into a single SCs.[116] Apart from the thiolate-protected co-crystals discussed above, a few co-crystals of dichalcogenide NCLs have also been reported.[131-134]

In close analogy to NCLs, the co-crystallization of different types of NCRs into SCs with an overwhelming structural diversity has been frequently achieved.[31] While most of these co-crystals exhibit closed-packed structures with direct atomic crystal counterparts,[135] some NCR co-crystals have even led to SC structures that do not have an atomic analogue, such as the $[PbSe]_6[CdSe]_{19}$ NCR SC[136] or the $[PbSe][Pd]_{13}$ SC, and ligand-ligand interactions are held responsible for this structural diversity.[137] Quasi-crystalline structures with five-, eight-, ten- and 12-fold rotations are likewise found in atomic crystals as well as NCR cocrystals. NCR co-crystals of semiconducting NCRs mixed with metallic NCRs have demonstrated electronic doping effects, and even more complex ternary NCR co-crystals have been reported, such as $[Fe_3O_4][PbS][CsPbBr_3]_3$ SCs with perovskite structure.[110] Recently Yu et al. reported SCs composed of CdS and PbS NCRs doped with atomically precise $Au_{12}$ NCLs resulting in an enhanced quantum yield.[138] This study may be viewed as an exciting example for future, truly interdisciplinary co-crystals of mixed NCL and NCR building blocks.

**Table 1.** Examples of noble-metal nanocluster co-crystals and their structures

| Crystallization method | Cluster 1 | Cluster 2 | Crystal System | Occupancy | Reference |
|---|---|---|---|---|---|
| Slow evaporation | $Ag_{16}(StBu)_8(CF_3CO_2)_7(CH_3CN)_3Cl$ | $Ag_{17}(StBu)_8(CF_3CO_2)_7(CH_3CN)_3Cl$ | Monoclinic | 2:1 | [139] |
| Slow evaporation | $Au_1Ag_{24}(SPhEt)_{18}$ | $Au_1Ag_{26}(SPhEt)_{18}(PPh_3)_6$ | Triclinic | 1:1 | [116] |
| Vapour diffusion | $Pt_1Ag_{28}(SAdm)_{20}$ | $Pt_1Ag_{28}(SAdm)_{18}(HO\text{-}SAdm)_2$ | Monoclinic | 1:1 | [129] |
| Layering method | $Ag_{40}(SPhMe_2)_{24}(PPh_3)$ | $Ag_{46}(SPhMe_2)_{24}(PPh_3)_8$ | Monoclinic | 1:1 | [127] |
| Slow evaporation | $Ag_{210}(SPh\text{-}4\text{-}iPr)_{71}(PPh_3)_5Cl$ | $Ag_{211}(SPh\text{-}4\text{-}iPr)_{71}(PPh_3)_6Cl$ | Triclinic | 1:1 | [128] |
| Layering method | $(AuAg)_{45}(SPhMe_2)_{27}(PPh_3)_6$ | $(AuAg)_{267}(SPhMe_2)_{80}$ | Hexagonal | 1:1 | [130] |

### 2.3 Charge-carrier transport in ensembles of nanocrystals or -clusters

Despite impressive advances in the surface chemistry of NCRs[140, 4, 141-143] and some examples with notable signs of band-like transport,[144, 145, 146] most NCR ensembles fall into the regime of "weak coupling", that is, temperature-activated transport.[26, 147, 148] This is particularly true for extended SCs of NCRs where a prerequisite for long-range structural order are relatively wide interparticle distances (> 1 nm) to prevent kinetic arrest during self-assembly. Transport in such materials is dominated by the interplay of three parameters: The average energetic fluctuations ($\Delta\alpha$), the charging energy ($E_C$) and the transfer integral ($\beta$). While $\Delta\alpha$ is determined by the chemical, structural or orientational disorder in the SC, $E_c$ is the energy barrier for injecting a charge carrier into an NCR or NCL of limited capacitance and $\beta = h/\tau$, with the hopping rate $\tau^{-1}$. If coupling is weak,

$$\beta_{weak} \approx 2\pi\epsilon \exp\left(-\frac{E_A}{k_B T}\right)$$

with the coupling energy $\epsilon$ and the activation barrier to transport $E_A$, which can be approximated as $E_A = E_C + \Delta\alpha$.[148] It is obvious that reducing the interparticle spacing in SCs will drastically enhance the hopping rate and, thus, the efficiency of charge carrier transport.

Within this framework, two regimes may be distinguished: 1) The Mott regime, where $E_C > \Delta\alpha$ and 2) the Anderson regime where $E_C < \Delta\alpha$ (**Figure 6**). In the Mott regime, If $\beta << E_C$ such that $E_C/(E_C+\beta) \to 1$, the so-called Hubbard gap opens and the SC exhibits the characteristics of a Mott insulator, following a temperature activated hopping mechanism. Only for $\beta > E_C$, a Mott insulator-metal transition occurs with the formation of delocalized states.[149] However if $\beta << \Delta\alpha$, Anderson localization occurs and the delocalization disappears. This highlights that increasing the chemical, structural or orientational order in SCs will have an advantageous effect on charge carrier transport.

Although SCs of epitaxially connected NCRs (e.g. with vanishing interparticle spacing) have become available[119, 150-155] and computational studies predict that strong coupling should be possible in these materials,[156, 157] their reproducible and large-scale synthesis remains a challenge.

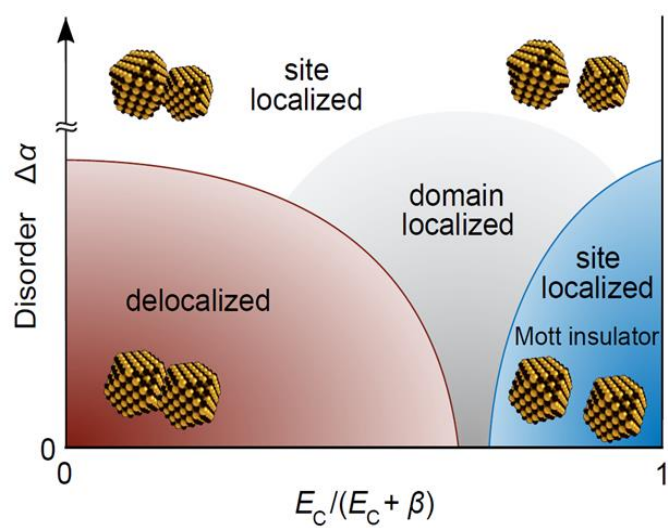

**Figure 6.** Schematic phase diagram of electronic states as a function of coupling strength $E_C(E_C + β)^{-1}$ and disorder $Δα$. The limits of strong and weak coupling correspond to $E_C(E_C + β)^{-1} → 0$ and $E_C(E_C + β)^{-1} → 1$, respectively. The two NCRs in each corner sketch the structural arrangement within the ensemble. Ligand spheres are omitted for clarity. Figure reproduced with permission from ref. 255.

Moreover, the conductivity of SCs of NCRs can substantially be enhanced by substitutional doping, where metallic NCRs are incorporated into semiconductor NCR superlattices to modulate carrier density and percolation pathways.[158]

Similar to NCRs, ensembles of NCLs exhibit temperature-activated charge-carrier transport as well.[159, 160, 161] Recent attempts to connect adjacent NCLs in SCs covalently have led to new materials with greatly improved mobilities,[120] but the interactions between neighbouring NCLs remained in the weak coupling regime. Clarke et al. investigated electric transport in films of $Au_{55}(PPh_3)_{12}Cl_6$ NCLs after electron-beam irradiation and observed characteristic features of Coulomb charging.[162, 163] Similar results were obtained for short, one-dimensional chains of these NCLs.[164] By crosslinking, the NCLs increase the electronic coupling, which thereby increases the conductivity.

**The effect of the interparticle distance:** The native ligand shell of colloidal NCRs restricts an effective interparticle coupling due to long chain lengths and wide band gaps, creating transport barriers of considerable width and height, which in turn results in insulating NCR ensembles.[5, 165]

A typical strategy for improving the electronic coupling between NCRs is to shorten interparticle distances by replacing these native ligands with shorter ligands, such as organic,[6, 141, 166, 167] or inorganic molecules.[5, 11, 140, 142, 168-170] In a seminal work by Talapin et al.[5] the conductivity of PbSe NCR solids was improved by ~ 10 orders of magnitude by the use of hydrazine as a small cross-linker, shortening the interparticle distance from ~ 1.1 nm to ~ 0.3 nm. Thus, ligand exchange has become a popular means to control interparticular distances and enhance the charge carrier transport within NCR materials.[6, 141, 171] With this, semiconductive NCR assemblies with charge carrier mobilities of up to $μ ≈ 1–24\ cm^2V^{−1}s^{−1}$ were achieved.[5, 6, 141, 168, 170, 172]

By cross-linking PbSe NCRs with bidentate alkanedithiol-ligands of various lengths, Liu et al. have proven – as theoretically predicted[149, 173] that the carrier mobility and thus the interparticle coupling increases exponentially with decreasing ligand length.[174] This exponential relation between the interparticular coupling and ligand length was further verified for semiconductive NCRs by other techniques.[167, 175] Larger Au NCRs feature the same behavior together with a metal-insulator transition for interparticle spacings of ~ 0.5–0.7 nm.[176, 177]

Such an exponential decay of the conductivity with increasing ligand lengths was also reported for small Au cluster with ~ 1 order of magnitude per Å.[178, 179] A recent, extensive treatment of electron transfer in thin films of $Au_{25}(SR)_{18}^0$ NCLs has been provided by Maran and co-workers.[180]

**The effect of the nature of the ligands:**

However, it is not only the interparticular distance that plays a key role in electronic coupling, but also the nature of the ligands themselves.[181, 182] In a remarkable study by Wessels et al. on Au NCRs, the use of π-conjugated ligands improved the conductivity and thus electronic coupling by one order of magnitude compared to ligands of the same length without π-conjugation.[181] Moreover, an exponential decay of the electronic coupling with increasing number of non-conjugated bonds within the ligand system was observed. Generally, the surface chemistry of NCRs is of significant importance.[20, 55] It has been demonstrated by Brown et al. that the position of the $1S_e$ and $1S_h$ energy levels of PbS NCRs can be shifted by up to 0.9 eV by functionalization of the surface with different ligands.[183] Besides these energetic effects, it has been revealed that for larger ligands the SC domain sizes became larger and the interdomain gaps were reduced, compared to shorter ligands.[184]

A similar ligand dependence of transport for NCL ensembles was evidenced by Galchenko et al. who reported the field-effect and photoconductivity of $Au_{25}$ NCL films.[159] $[Au_{25}(PPh_3)_{10}(SC_2H_4Ph)_5X_2]^{2+}$ NCLs exhibited an n-type field effect and pronounced photoconductivity, which are clear signs of a semiconductor. In contrast, $[Au_{25}(SC_2H_4Ph)_{18}]^{1−}$ NCLs lacked such behaviour and were found poorly conductive.

## 3 Emerging properties

### 3.1 Emerging optical properties

#### 3.1.1. Nanoclusters:

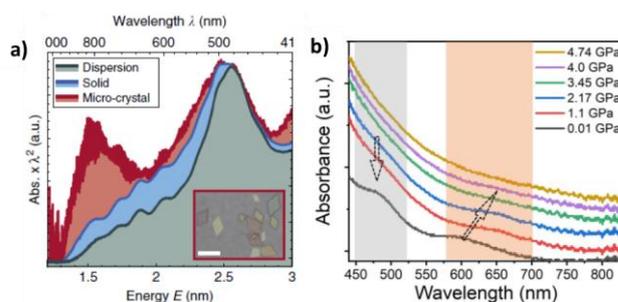

**Figure 7. a)** Absorption spectra of $Au_{32}$ NCLs in dispersion, thin-film, and supercrystal. Figure reproduced with permission from ref. 160. **b)** Pressure-dependent absorption of $Au_{28}$ NCLs. Figure reproduced with permission from ref. 185.

In 2015, Lina et al. reported that upon crystallization, the absorption spectrum of $Ag_{29}$ NCL SCs is broadened and red-shifted.[98] Our group has investigated an enhanced and redshifted absorption near the band edge transition in SCs of $Au_{32}$ NCLs compared to their absorption in solution or in a glassy thin film (**Figure 7a**).[160] While one likely explanation for this phenomenon is enhanced electronic coupling due to the close proximity of adjacent NCLs, an additional factor may be residual stress in the SCs, which is often introduced during drying.[118] In line with this, Li et al. observed a red-shift in a series of face-centered cubic (fcc) and bi-tetrahedral NCLs ($Ag_{28}Pt_1$, $Au_{21}$, $Au_{28}$, $Au_{24}$, and $Au_{14}Cd$) upon applying uniaxial pressure using a diamond anvil cell (**Figure 7b**).[185] This was ascribed to a stress-induced electron delocalization of the core electrons to the ligands.

Another emergent optical property, which is often found in SCs of NCLs, is crystallization induced emission enhancement (CIEE).[38,43] Chen et al. first reported CIEE in SCs of bimetallic $Au_4Ag_9$ NCLs, co-protected by thiolate and phosphine ligands (**Figure 8a**).[43] The highly ordered SCs displayed strong luminescence at 695 nm, whereas the solution exhibited no emission. The enhanced emission was attributed to restricted vibrations and rotations in the crystalline state. Structural investigations revealed that the tri-blade fan-like configuration of the NCL framework and the C−H···π interactions resulted in structural restrictions. Similarly, Khatun et al. reported a 12-fold enhancement of the emission in $Ag_{22}$ NCL SCs (**Figure 8b**).[256] Detailed structural analysis of the SCs revealed that the intermolecular non-covalent interactions of the ligands, such as CH···π and π···π interactions, cause the CIEE. Since then, a growing number of more examples for CIEE in NCLs were reported, which followed a similar mechanism.[99, 186, 187]

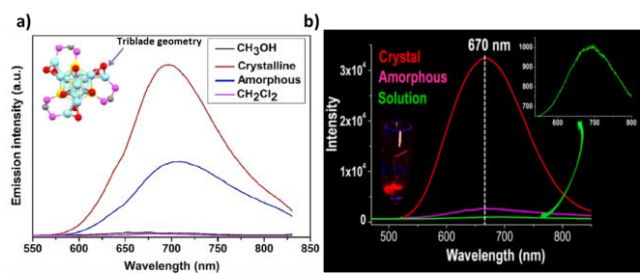

**Figure 8. a)** Photoluminescence of the crystalline (red), solution (magenta for $CH_2Cl_2$, black for methanol), and amorphous (blue) states of $Au_4Ag_{13}$ NCLs. Figure reproduced with permission from ref. 43. **b)** Photoluminescence of $Ag_{22}$ NCLs in the crystalline (red), amorphous (pink), or solution (green) state. Figure reproduced with permission from ref. 256.

In this context, a pressure-dependent study of a series of fcc and bi-tetrahedral ($Ag_{28}Pt_1$, $Au_{21}$, $Au_{28}$, $Au_{24}$ and, $Au_{14}Cd$) NCL SCs showed that up to 200-fold fluorescence enhancements are possible under stress in these materials. This was attributed to the suppression of non-radiative vibrations, enhanced near band-edge transition dipoles and the absence of excited-state structural distortion.[185]

### 3.1.2 Nanocrystals:

One of the first pieces of evidence for collective electronic properties in NCR ensembles was found spectroscopically. A comparison of NCRs (semiconductive and metallic) in solution and in films has shown that additional optical transitions or a broadening and redshift occur in densely packed films. This was interpreted as an evolution from individual localized to collective electron states delocalized within at least a few NCRs.[188-193] Meanwhile, many more studies have demonstrated the effect of collective emerging optical properties of self-assembled NCRs. One example for this is the coupling of plasmonic modes within highly ordered SCs of metallic NCRs.[194] Here, the electrons of the NCRs are localized, while collective plasmonic states appear.[75] The collective coupling strongly depends on the particle size, interparticle spacing and surrounding, creating an electric field enhancement due to the localized surface plasmon resonance (LSPR) at the interparticle gaps.[83] A remarkable computational study showed that the properties of the collective plasmon modes are affected by the unit cell symmetry of the SC of metal NCRs.[195] Schulz et al. have demonstrated the experimental realization of highly ordered SCs of >20 nm Au NCRs with hexagonal symmetry. The high degree of periodicity induced novel well-defined collective plasmon-polariton modes.[75, 196]

Bian et al. have prepared large-scale SCs with an hcp lattice of ~4.4 nm Au NCRs. The plasmonic coupling for these highly ordered SCs was larger than for amorphous films of the same NCRs. The plasmonic coupling appeared to be facet-dependent, evidenced by the finding that it was largest for the facet with the largest number of nearest neighbours and highest degree of symmetry.[83]

Another example for coupled dipoles in highly periodic SCs is the superfluorescence of caesium lead halide ($CsPbX_3$, X = Cl, Br, I) perovskite NCRs, self-assembled into SCs. This phenomenon is caused by the excitation of an ensemble of individual dipole emitters and results in a macroscopic quantum state, causing collective emission.[197] This superfluorescence – only observed at cryogenic temperatures, yet – is a key property to create spectrally ultra-pure laser sources or highly efficient light-harvesting systems. Quantum chemical simulations have shown that structural order in the SCs and its effect on the thermal decoherence play a crucial role in the efficiency of the superfluorescence.[198] Thus, controlling structural defects in these SCs, such as isolated NCR vacancies which have been studied by electron microscopy, is key for efficient future lighting application.[199] We note that there are also other explanations for the apparent emergent optical properties of SCs of highly monodisperse $CsPbX_3$ perovskite NRCs. A confocal fluorescence microscopy study on $CsPb(I_{(1-x)}Br_x)_3$ NCR SCs revealed a spectral blueshift near the edges of the SC under intense optical excitation.[200] This was attributed to a gradual contraction of the atomic lattice and a photoinduced loss in iodine, starting first from the edges of the SC.

In a similar context, Shevchenko et al. investigated the optical properties of SCs of monodisperse CdSe NCRs.[201] It was shown that the fluorescence signal from the edges of the 50 μm sized SCs is spectrally blue-shifted with slightly increased fluorescence lifetimes. This was attributed to the coexistence of two types of CdSe NCRs due to partial oxidation of the CdSe NCRs near the edges.

We note that, in principle, the photonic bandgap found in photonic crystals (e.g. from silica particles) is another excellent example for emergent optical properties in NCR SCs.[202] We do not cover this topic here since we have deliberately limited ourselves to building blocks with a diameter of 20 nm or less (see section 2.1). Since photonic crystals and their emergent optical properties, such as a negative index of refraction,[203] require periodicities of roughly half the wavelength of the incoming light, analogous effects exhibited by the NCRs covered here would only be applicable to photons of ~30 eV, which are technologically of little relevance.[204]

## 3.2. Emerging transport properties

### 3.2.1 Nanoclusters:

#### The effect of structural order

Li et al. demonstrated that electric transport in of $Au_{21}(SR)_{12}(PCP)_2^+$ NCL SCs depends on the structure of the SC, which can be controlled by the counter ion (**Figure 9a+b**) PCP: bis(diphenylphosphinomethane).[114] With $[AgCl_2]^-$ as the counterion, a triclinic structure was obtained in which the NCLs are linearly assembled along the diagonal of the {100} plane, whereas for $[Cl]^-$ as the counterion, a monoclinic lattice with a linear assembly of the NCLs along the diagonal of the {010} plane was obtained. The monoclinic $[Au_{21}(SR)_{12}(PCP)_2]^+[Cl]^-$ SCs exhibited a two orders of magnitude higher conductivity (σ ~ 2.38×10$^{-6}$ Sm$^{-1}$) than the triclinic $[Au_{21}(SR)_{12}(PCP)_2]^+[AgCl_2]^-$ SCs (σ~1.44×10$^{-8}$ Sm$^{-1}$). This was explained with the parallel-displaced π-stacking of the PCP-phenyl rings in the case of the monoclinic SCs, which presumably reduces the barrier to hopping transport.

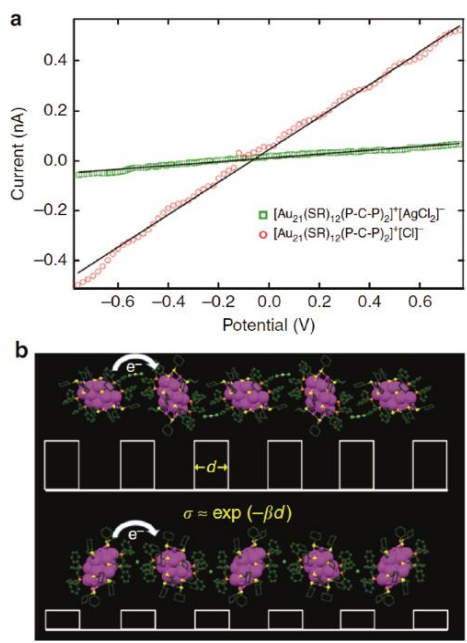

**Figure 9. a)** Room temperature conductivity of $[Au_{21}(SR)_{12}(PCP)_2]^+[AgCl_2]^-$ (green) and $[Au_{21}(SR)_{12}(PCP)_2]^+[Cl]^-$ (red) NCL SCs, respectively. (PCP: bis(diphenylphosphinomethane)) **b)** Schematic representation of the hopping mechanism in the two $Au_{21}$ NCL SCs, where the different structure results in different tunnelling barrier heights (white squares). Figure reproduced with permission from ref. 114.

Maman et al. reported an approx. three orders of magnitude conductivity increase upon a reversible structural transformation of $Ni_6(PET)_{12}$ NCL SCs (PET = phenylethanethiol) upon heating to 130 °C. While both SC types had the same monoclinic space group, their structures differed in the non-covalent ligand-ligand interactions, specifically the S···S distances between neighbouring PET moieties. For the more conductive polymorph, a shorter S···S distance was detected (3.78 Å vs. 4.00 Å).[205]

Li et al. have designed SCs of penicillamine-protected $Ag_{14}$ NCLs and discussed the feasibility of proton conductivity across the $-NH_3^+$ and $-COO^-$ ion pairs of the penicillamine ligands.[206] The $Ag_{14}$ NCL displayed a proton conductivity of $3.06 \times 10^{-5}$ Scm$^{-1}$ at 30 °C, which increased to $1.94 \times 10^{-4}$ Scm$^{-1}$ when the temperature was raised to 60 °C.

**Effect of interconnectivity**

Yuan et al. investigated anisotropic electric transport in SCs of 1-ethynyladamantane-protected $(AuAg)_{34}$ NCLs (**Figure 10a**).[120] The monoclinic SCs exhibited Ag-Au-Ag bonds between adjacent NCLs to form one-dimensional polymer chains of connected $(AuAg)_{34}$ kernels. Along the crystallographic direction of this chain, the electrical conductivity was 1800 times higher ($1.49 \times 10^{-5}$ Sm$^{-1}$) than in the cross directions. A hole mobility of 0.02 cm$^2$V$^{-1}$s$^{-1}$ and an ON/OFF ratio of 4000 indicated ample p-type semiconducting behaviour. These anisotropic SCs of NCLs may be viewed as analogues to epitaxially connected NCRs obtained by oriented attachment.[119]

Following a similar idea, Tang et al. have described trigonal SCs of $K_4Cu_2Ge_3Se_9(H_2O)$ NCLs, which form one-dimensional chains. The NCLs consisted of tetrahedal GeSe clusters, which are interconnected by $Cu^+$ atoms, and exhibited a unique nanotubular structure. In the direction of the main axis of the nanotube, the electrical conductivity was $7.60 \times 10^{-6}$ S cm$^{-1}$ at 40 °C.[207]

### 3.2.2. Nanocrystals:

**The effect of structural order**

A study on Au NCR SCs has shown that the threshold voltage to overcome the Coulomb blockade to transport strongly depends on the amount of interparticle voids, which emphasizes the need for low defect densities.[208] The inherent energetic and spatial disorder in SCs of NCRs (see section 2.1.2) can lead to a transition from delocalized to localized states if the disorder exceeds a critical value.[209] It was shown spectroscopically that a size distribution-based disorder adversely affects the exciton diffusion rate in NCR assemblies.[210, 211]

A computational study established that the critical disorder depends strongly on the coordination number of the NCRs inside an SC and, thus, on its unit cell.[209] Briefly, more nearest neighbours introduce, on average, more coupling to each NC and decrease the overall fluctuation of coupling energy. While no effect of long-range structural order could be found, three-dimensional SCs appeared to be much better coupled than two-dimensional assemblies, indicating that dimensionality plays an important role in localization.

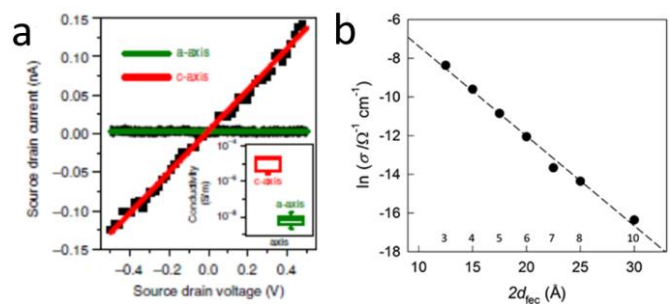

**Figure 10. a)** I–V plot of polymeric $(AuAg)_{34}$ NC SCs measured along the a-axis and c-axis, respectively. Figure reproduced with permission from ref. 120. **b)** Variation of film conductivity as a function of twice the length $d_{fec}$. Figure reproduced with permission from ref. 180.

Temperature-dependent transport measurements of large Ag NCR SCs revealed that the temperature at which the mechanism transits from a variable-range hopping to simple activated hopping depends on the size dispersion of the NCRs.[212] However, attempts to prove the detrimental effect of size-induced energetic disorder in semiconducting NCRs were unsuccessful, as Liu et al. have shown that the carrier mobility is essentially independent of the size distribution,[174] although an effect was computationally predicted.[156] However, it could be argued that interparticle coupling in the investigated ensembles was too low to observe the effect.

Nair et al. reported substantially higher conductivities for polycrystalline ensembles of large Au NCRs compared to amorphous films of the same NCRs.[213] A similar effect was observed for hydrazine-treated PbS NCRs, where the conductivity and mobility (μ ≈ $10^{-11}$ to $10^{-8}$ cm$^2$ V$^{-1}$ s$^{-1}$) of polycrystalline films with grain sizes of ≤ 100 nm exceeded that of glassy films.[214] Recently, it was shown by applying time-resolved spectroscopy and X-ray scattering that the charge carrier hopping rate in PbS NCR SCs was enhanced by improving the degree of structural order and NCR alignment induced by thermal treatment.[215]

Kaushik et al. have sketched an appealing idea that should apply to all ordered SCs in the weak coupling regime:[216] due to the strong dependence of coupling on the interparticle distance (see section 2.3), transport in SCs should be most efficient in the crystallographic direction of the nearest-neighbour distance. Specifically, their computational analysis showed that both, the electron and hole mobilities, along the nearest-neighbour direction are 4–6 orders of magnitude larger compared to the direction of second-nearest neighbors for bcc and fcc lattices of PbSe NCRs.

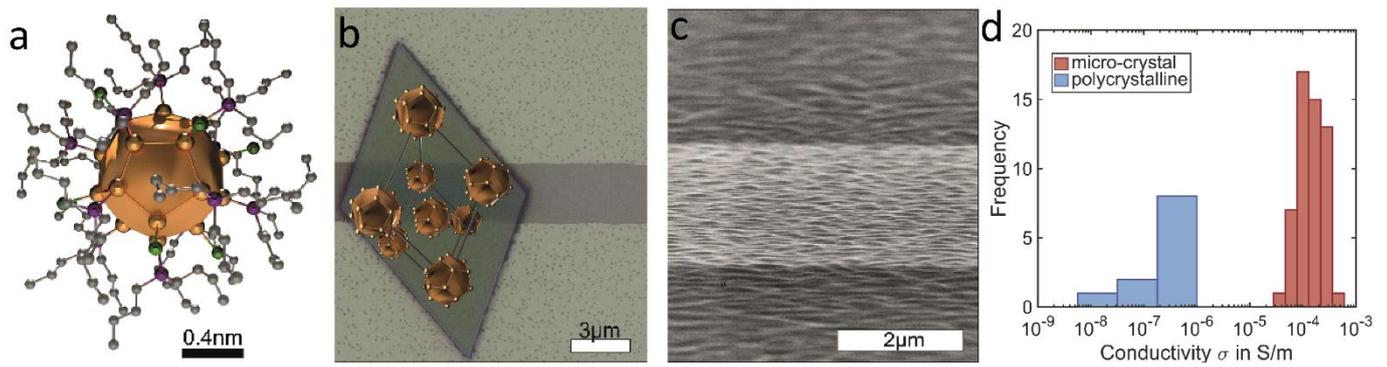

**Figure 11. a)** Structure of an $Au_{32}(^nBu_3P)_{12}Cl_8$ NCL. The different colours represent the Au- (gold), Cl- (green), P- (purple) and C- (grey) atoms, while hydrogens are omitted for clarity. **b)** Electron micrograph of an individual SC deposited on two horizontal Au-electrodes on Si/SiO$_x$, including a schematic drawing of the triclinic unit cell of the SC. **c)** Electron micrograph of a thin, polycrystalline film of the same NCLs (Au electrodes in dark, channel gap in bright). **d)** Conductivity of 54 individual SCs (red) and 19 polycrystalline thin films (blue). Figure adapted with permission from ref. 160.

**Effect of orientational order**

Computational studies have shown that the electronic coupling between NCRs strongly depends on their orientation to each other.[188] The same study has revealed a strong effect of the NCR shape on the electronic coupling, as the coupling energy of cubic NCRs exceeds that of spherical NCRs, attributed to an improved degree of facet alignment. The computational study by Kaushik et al. mentioned above has observed that the electronic coupling between two adjacent PbSe NCRs differs for electrons and holes, and is further facet-dependent: electron coupling is generally stronger on smaller facets, whereas hole coupling is stronger on larger facets.[216] Consequently, for cuboctahedral NCRs where the {100} facets are larger than the {111} facets, electron coupling preferentially extends along the [111] direction, whereas the hole coupling is stronger along [100]. Based on these computational considerations, an experimentally realized orientational alignment of the individual NCRs within an SC seems beneficial.

The occurrence of preferred coupling directions implies that charge transport in long-range ordered NCR SLs might be anisotropic. This hypothesis is further strengthened by an experimental study on individual ~ 200 nm sized PbS (nano)crystals, which revealed a facet-dependent electrical conductivity.[217] High conductances were found through the {100} and {110} facets, whereas the {111} facet remained insulating. Moreover, a computational study by Yazdani et al.[218] on small-polaron hopping in PbS NCR SCs concludes a similar facet-dependent charge transport anisotropy.

Oriented attachment of adjacent NCRs (mainly lead chalcogenides) has resulted in honeycomb or square lattice structures, where comparable improvements of the interparticular coupling and charge carrier mobilities in the range of $\mu \approx 0.2-13$ cm$^2$ V$^{-1}$ s$^{-1}$ can be obtained.[119, 151, 152, 219-222] 2D PbSe honeycomb structures have successfully been cation exchanged to CdSe honeycomb structures, which are predicted to feature novel electronic band structures of Dirac bands combined with a strong spin-orbit coupling.[119, 223]

# 4 A personal view of the field

## 4.1 The effect of long-range structural order

We developed a microcontact printing technique to fabricate micrometer-sized channels of PbS NCR SCs with long-range order and dimensions approaching the size of typical single-crystalline domains of only a few µm$^2$.[224] Comparing the electrical conductivity in such SC microchannels with that of glassy domains over large scales, we found transport to be orders of magnitude more efficient in the SC channels with long-range structural order.[224] We used X-ray nanodiffraction to determine the local structure of the PbS NCR SCs within the microchannels and correlated it with the electric transport properties of the individual channels. We identified two types of structures for all examined SCs: a polycrystalline body-centered cubic (bcc) structure or a monocrystalline random hexagonal closed-packed (rhcp) structure. We made two key observations: 1) The NCRs in the polycrystalline bcc SC exhibited a narrow range of nearest-neighbor distances, which were significantly shorter than those in the monocrystalline rhcp SCs. This emphasizes the necessity of a minimum interparticle distance to maintain large structural coherence and a high degree of long-range order. 2) Although temperature-activated hopping should be more efficient in the polycrystalline bcc SCs with shorter interparticle distances (see section 2.3), we found that the conductivities of the monocrystalline rhcp SCs could be as high as those of polycrystalline SCs. We consider this an indication that the degree of crystallinity can at least partially compensate for the deleterious effect of the much larger interparticle distance, such that structural order has a significant effect on the conductivity in NCR SCs.[94]

In a similar context, we have investigated the effect of structural order on the electronic properties of $Au_{32}(^nBu_3P)_{12}Cl_8$ NCLs (**Figure 11**).[160] To this end, we prepared two types of materials with the same NCL building blocks: Via spin-coating, we obtained thin films with glass-like ensembles of NCLs. Upon slow self-assembly at the acetonitrile/air interface, we generated rhombohedral crystals of up to 30 µm in length, which appeared mostly single-crystalline. For these SCs with long-range structural order, we observed a two-orders of magnitude increase of the electric conductivity ($\sigma = 1.56 \times 10^{-4}$ Sm$^{-1}$) compared to glassy ensembles of the same NCLs ($\sigma \approx 1 \times 10^{-6}$ Sm$^{-1}$). We used temperature-dependent electric transport measurements to derive the activation barrier for carrier hopping in these two types of samples via a simple Arrhenius model. For the glassy ensemble, we found significant contributions to this barrier from both, the charging energy as well as energetic disorder (see section 2.3). In contrast, for the single-crystalline SCs, the activation energy was essentially equal to the charging energy, implying that the SCs exhibited a vanishing energetic disorder, which we held responsible for the two-orders of magnitude increase in the conductivity. This result emphasises the effect of long-range structural order on transport through NCL ensembles (**Figure 12**), which is similar to the effect observed in atomic crystals, e.g. the case of crystalline vs. amorphous silicon.

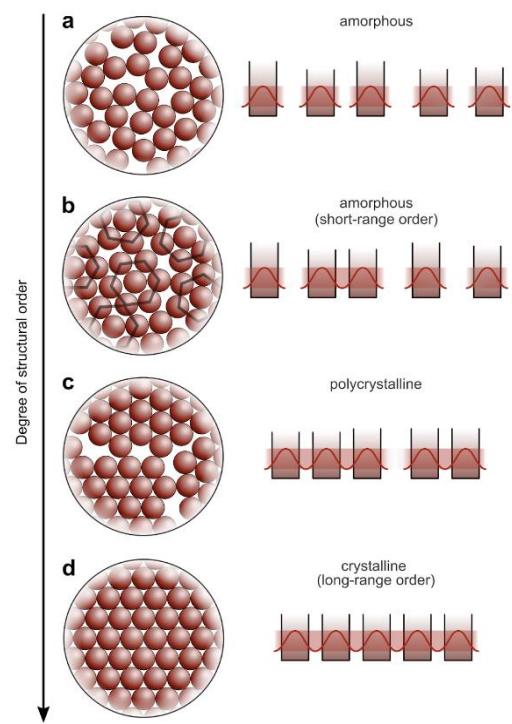

**Figure 12**. Different degrees of structural order (**left**) and their effects on interparticle coupling (**right**): **a)** fully amorphous ensemble of hard spheres, **b)** glassy ensemble with short-range order, **c)** polycrystalline ensemble with ordered domains, and **d)** supercrystal with long-range order. Electronic coupling, represented by red bands, increases towards the bottom, but remains weak as indicated by the localized wave functions. Figure adapted with permission from ref. 255.

### 4.2 The emergence of transport anisotropy

In the study of PbS NCRs mentioned above, we found strong experimental evidence of anisotropic charge transport within ordered SCs.[94] For monocrystalline rhcp SCs oriented along the [0001] direction, the influence of the in- plane SC orientation with respect to the electric field direction on the conductivity was investigated. Compared to otherwise identical SCs where the nearest-neighbour distance was oriented at a non-zero angle with respect to the electric field, we observed a 40–50% higher conductivity for SCs where the nearest-neighbour direction was oriented parallel to the electric field (**Figure 13**). This iso-orientation results in the shortest hopping distance and the fewest number of hops across the transport channel. For all other orientations, the charge carriers must either hop over larger distances or take more hops across the nearest-neighbour pathways in a zig-zag fashion (see **Figure 13d**). This result should be compared to the computational study by Kaushik et al. in section 3.2.2., who predicted precisely this behaviour.[216] Such anisotropic charge transport is the result of the most efficient charge carrier hopping along the shortest interparticle distance and, thus, is assumed to be a general feature of weakly coupled NCR SCs. Consequently, one can expect for simple cubic, fcc or bcc SCs that their favoured directions for charge transport will be the corresponding nearest neighbour directions, which are the <100>, <110> or <111>, respectively. We further observed that the orientational order of the NCRs might be an additional source for the occurrence of a favoured charge transport direction. Caused by the mesocrystallinity of the rhcp SCs, the [110] directions of all NCRs are iso-oriented with the nearest-neighbour distance.

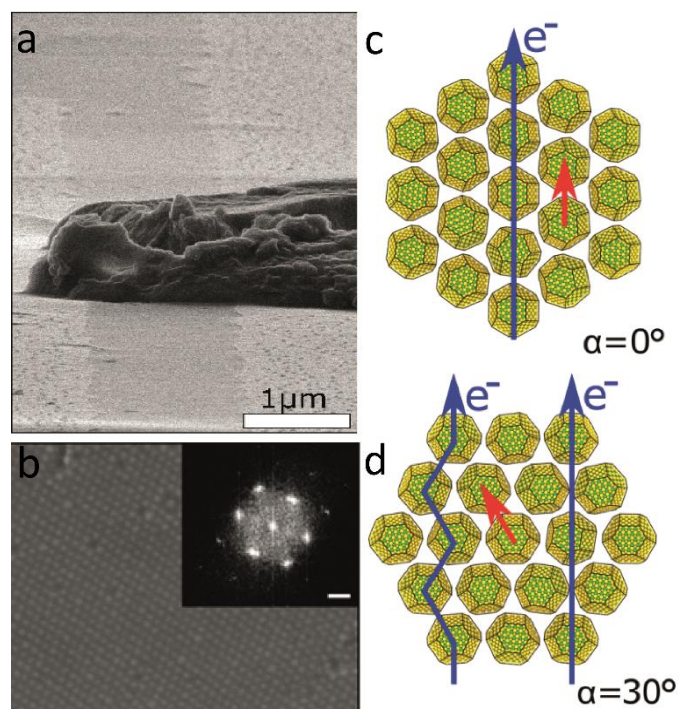

**Figure 13. a)** Electron micrograph of a PbS NCR SC bridging the channel between two Au electrodes. **b)** Top view of the SC and its Fourier transform in the inset. **c)** Schematics of the SC oriented along the (0001) direction with **c)** α = 0° and **d)** α = 30°. The direction of the electric field and the nearest-neighbour direction are indicated with blue and red arrows, respectively. For an in-plane offset (α = 30°), the larger hopping distance or the zig-zag path are detrimental to charge transport. Ligand spheres of NCRs are omitted for clarity. Adapted from ref. 94.

The existence of anisotropic charge transport is already known for organic semiconductors[225] or metal-organic frameworks,[226] where charge transport differs for different crystallographic directions. Additionally, in conjugated polymers a strong correlation between structural order and charge transport efficiency has been found.[227]

### 4.3 The appearance of surface defects and their correlation with the optical properties

We have studied the effect of surface defects in SCs of caesium lead halide perovskite NCRs on their spatially resolved fluorescence properties (**Figure 14**).[228] We self-assembled the SCs by simple solvent evaporation (see **Figure 5a**, section 2.2) and analysed their structure by nanodiffraction with a 400 nm x 400 nm X-ray beam as introduced in section 2.2. By scanning individual SCs with this beam, we observed near-single crystalline SCs with typical grain sizes of 10 µm² **Figure 14e**). We used diffraction-limited confocal laser scanning microscopy (**Figure 14b**) to measure the local photoluminescence (**Figure 14 c**) and its lifetime (**Figure 14d**) across individual SCs with a spatial resolution of 200 nm. We found a decrease in interparticle distance (**Figure 14 f**), a loss in structural coherence and orientational order (**Figure 14 g**) as well as compressive strain near their surfaces (**Figure 14 h**). These structural distortions were strongly correlated with a blue-shifted fluorescence and decreased radiative lifetimes. We interpreted these findings with a reduced stability of the excited state of the emitting NCRs at the edges of an SC. With regard to testing the hypothesis of an analogy between atomic crystals and SCs, it is expedient to compare these results with the correlation between structural distortions and the fluorescence properties in atomic crystals. A good model system in this respect are monolayers of transition metal dichalcogenides, such as $WS_2$, where essentially all atoms are surface atoms. Conductive atomic force microscopy, Raman spectroscopy, scanning photoelectron microscopy and nanoscale photoluminescence spectroscopy have shown that the edges of these atomic crystals exhibit photoluminescence around 25 times stronger than at the center, which was attributed to a larger population of charge carriers at the edges. In general, a strongly inverse relationship between photoluminescence intensity and defect density was observed.[229, 230, 231-233]

Although we did not investigate potential superfluorescence from these SCs (see section 3.2.1.), we concluded that for this emergent optical property to appear efficiently, it would be mandatory to mitigate surface strain and orientational disordered as much as possible, e.g. by optimizing the self-assembly using some of the concepts introduced in section 2.2.

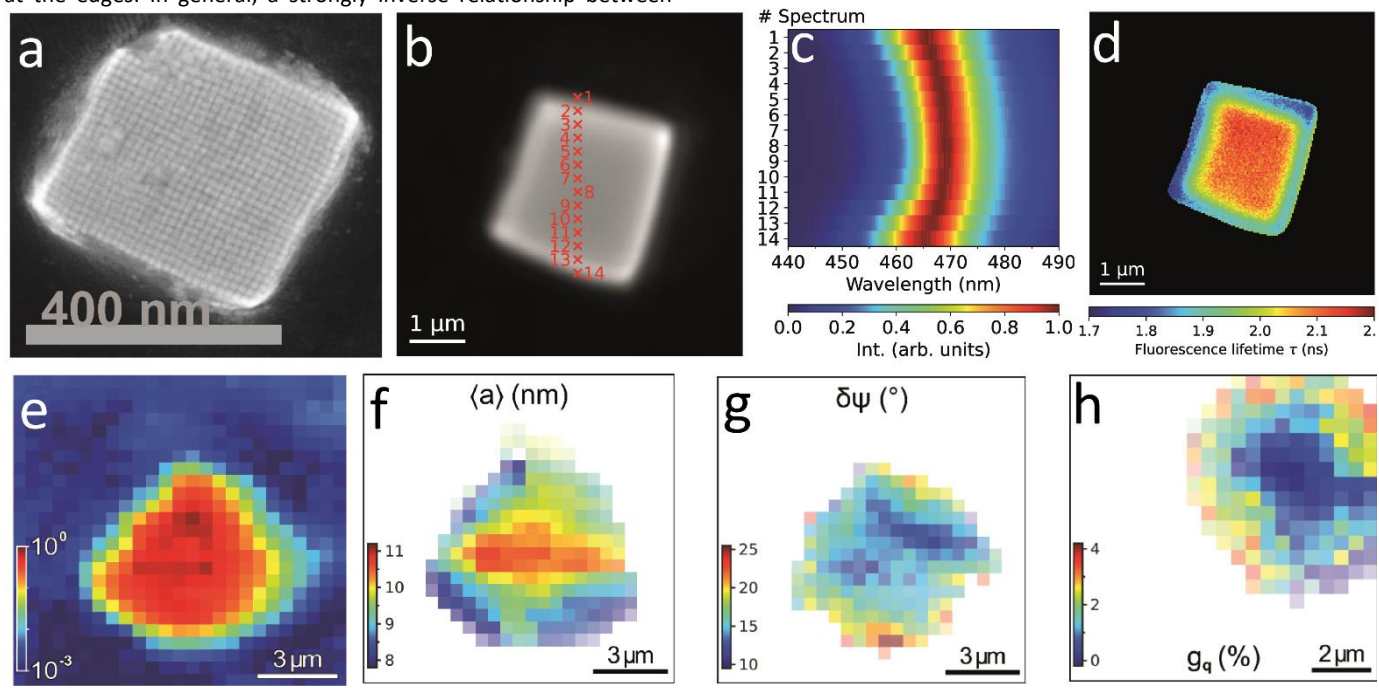

**Figure 14. a)** Electron and **b)** optical micrograph of a $CsPbBr_2Cl$ NCR SC. Positions of the measured photoluminescence spectra are indicated. **c)** The corresponding normalized fluorescence spectra. **d)** Spatially resolved fluorescence lifetime. **e)** X-ray micrograph of a similar $CsPbBr_2Cl$ NCR SC with a step size of 500 nm. **f)** Spatially resolved, average unit cell parameter of this SC, indicating a compression toward the surface. **g)** Spatially resolved angular disorder of the NCRs within the SC. **h)** Spatially resolved distortion of the atomic lattices of the NCRs within the SC. Figure reproduced with permission from ref. 228

### 4.4 Correlation between the type of unit cell and the mechanical properties

Quantum mechanical simulations of $Na_4Ag_{44}$ NCL SCs have predicted anomalous pressure softening and a compression-induced transition to a soft-solid phase due to ligand-ligand interactions.[108] Sugi et al. have experimentally probed the mechanical response of cubic $Ag_{29}(BDT)_{12}(TPP)_4$ SCs (abbreviated here as "$Ag_{29}$ C"; BDT = 1,3-benzenedithiol, TPP = triphenylphosphine) and indeed found a ligand interaction-induced softness of the SCs.[234] In a follow-up study on these dithiol-stabilised $Ag_{29}$ NCLs of cubic or trigonal polymorphs ($Ag_{29}$ T), monothiol-stabilised, trigonal $Ag_{46}$ ($Ag_{46}$ T) and their monoclinic $Ag_{40}$ co-crystals ($Ag_{40/46}$ M), differences in the Young's modulus ($E_r$) and hardness ($H$) were investigated (**Figures 15a, 15b**).[235] A central finding was the higher $E_r$ and $H$ for the dithiol-protected NCLs compared to the monothiol-counterparts, indicating the important role of structural rigidity due to the bidentate ligands. The second finding was the significant dependence of $E_r$ and $H$ on the SC structure with otherwise identical ligands for the two $Ag_{29}$ NCL polymorphs. The larger $E_r$ and $H$ values of the trigonal polymorph were explained with the larger number of interactions between BDT ligands, which are possible in the SC structure. Similarly, the slightly higher $E_r$ and $H$ values for the $Ag_{40/46}$ M co-crystals compared to $Ag_{46}$ T were ascribed to a larger number of interactions between the monothiol ligands that manifest in this SC structure. This study has demonstrated that different mechanical properties emerge from similar or even the same NCLs upon crystallization into SCs with different unit cells.

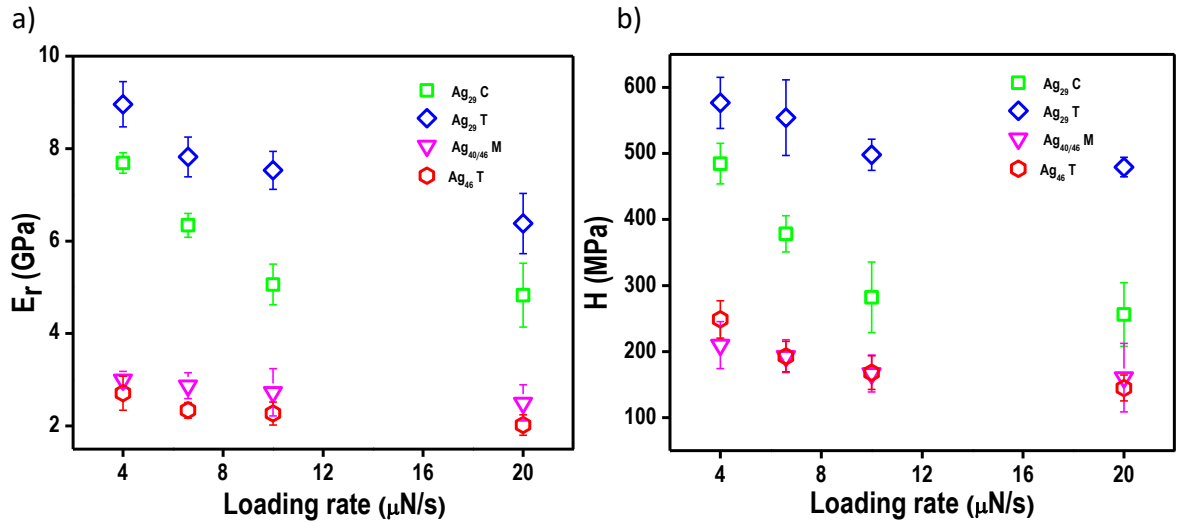

**Figure 15. a)** and **b)** Variation of Young's modulus and hardness as a function of loading rate of $Ag_{29}$ C (green), $Ag_{29}$ T (blue), $Ag_{40/46}$ M (pink), and $Ag_{46}$ T (red). Redrawn from ref. 235.

## 5 Outlook

### 5.1 Challenges

With regard to emergent electric transport properties in SCs, a persistent drawback is the absence of strong coupling and band-like transport in these materials. As outlined in section 2.3, chemical means, in particular ligand exchange, have been developed to fabricate NCR ensembles in which the typical attributes of band-like transport are observed. However, the required increase in coupling for this to occur has only been possible by largely reducing the interparticle spacing, which typically introduces kinetic arrest and prevents long-range, crystalline order (section 2.2). Within the nanocrystal community, a common hypothesis is that, if synthetic methods excelled further in narrowing the size-distribution of the nanocrystals, strong coupling in SCs may eventually become possible. However, no NCR will ever be more precise than, for instance, a $Au_{32}(^nBu_3P)_{12}Cl_8$ cluster, and the fact that SCs of such clusters are also weakly coupled sheds some doubts on this hypothesis.

As demonstrated in section 4.3, even highly ordered SCs are subject to surface tension, which one may compare to some degree to surface reconstructions in atomic crystals. This tension increases the probability of structural defects to occur with potentially detrimental effects on the collective electrical, mechanical, or optical behaviour of the SC. With regard to electric transport, surface defects are likely to increase the contact resistance, while for emergent optical properties, such as superfluorescence, it may prevent a uniform, collective response entirely.

A specific drawback of many NCLs is their limited photostability, at least in the solid state. For all NCLs summarised in section 2.1.1., there are examples for bulk crystals of the pure metals, e.g. fcc gold. The packing density of the metals in such crystals is often several times larger than in the corresponding NCLs, and this renders recrystallization thermodynamically favourable. Under optical excitation, the kinetic barrier against recrystallization, enabled by the ligands of the cluster, can be overcome and the NCLs irreversibly transform into the bulk metal analogues. Therefore, the emergent optical properties of NCL SCs are best exploited by low-intensity or off-resonant applications.

For truly collective phenomena in SCs to occur, it is indispensable that the building blocks be as uniform as possible. A demonstration of this important paradigm are the difficulties of realising a strong quantum confined Stark effect in NCR ensembles.[236-238] In principle, this effect can be very strong in SCs built from zero-dimensional objects.[239] However, if the building blocks exhibit only slight non-uniformities, spectral broadening will occur with deleterious effects on the efficiency of electro-optical modulation.[240]

When tuning the SC unit cell from one type to another, a set of synthetic parameters needs to be altered that also affect carrier transport. For example, PbS NCRs passivated by a ligand shell of oleic acid behave as hard spheres with short-range attraction and form a close-packed SC with fcc symmetry, together with a random orientation of the NCR cores. In contrast, the same NCRs with an incomplete passivating ligand shell form a bcc SC, where the NCRs are iso-oriented.[241] This is attributed to a preferred ligand coverage on the {111} NCR facets, resulting in a better ligand interaction between adjacent NCRs in the bcc lattice together with an increased NCR core interaction.[61, 102, 241] Consequently, detailing the pure effect of the SC unit cell of the same NCRs on electric transport requires some means to account for the additional effect of different ligand densities.

### 5.2 Possible solutions

Oriented attachment by controlled necking of adjacent NCRs into square- or honeycomb superlattices (section 2.2) has emerged as a promising combination of strong electron coupling and long-range structural order. DC mobilities as high as 13 cm$^2$/Vs have been demonstrated, which could be seen as a sign of strong coupling,[152] such that these materials bear the potential to enable the first experimental demonstration of the theoretical prediction of miniband formation in SCs.[242] For this to happen, further improvements of the reproducibility and fidelity of these complex materials would be necessary. A similar approach for NCLs is the idea of covalent linkage between individual shell atoms of adjacent NCLs (section 3.2.2.). This concept takes full advantage of the structural uniformity of NCLs and the feasibility of reproducible, atomically precise chemical interactions between the NCLs, which are not possible to this degree with NCRs. However, the mobilities of 0.02 cm$^2$/Vs indicate that further research into this material class is necessary. An alternative approach towards crystalline SCs with strong coupling is exchange of the building blocks with organic pi-systems and tailored energy levels to achieve resonant alignment between particles and ligands.[165, 243] This concept aims at reducing the hopping barrier height rather than its length (see section 2.3) such that much larger interparticle distances can be tolerated for strong coupling and long-range order can prevail. Although certainly not explored to its limits, this concept has achieved mobilities of, at best, 0.2 cm²/Vs, indicating that strong coupling has not been realised so far.[244]

We believe that the challenges associated with surface stress in SCs as well as the difficulties of changing only the SC unit cell without modifying other system parameters are already addressable with the tools described in section 2.2, provided more research is conducted in this direction. The input by computational chemistry has delivered important stimuli for the experimental advances that have already been achieved, and we deem continued exchange between theory and experiment particularly rewarding in this field. In terms of the necessity to fabricate SCs with highly uniform building blocks to create strong collective effects, atomically precise NCLs are arguably advantageous over NCRs. We note that "self-cleaning or -purification" effects as a result of self-assembly can mitigate this drawback of NCRs to some extent, and ensembles with very high uniformity are obtained if applied effectively.[245, 246] Our results in section 4.1 have shown that even single-crystalline SCs of atomically precise NCLs do not necessarily exhibit band-like transport, presumably due to the large charging energy in these small building blocks with limited capacitance. Thus, future NCL SCs should utilise larger NCLs (with higher capacitance) and/or reduce the interparticle distance to escape the Mott-insulator regime (see **Figure 6**).

### 5.3 Opportunities

While currently mostly of relevance for basic research, there is substantial potential for multinary SCs of mixed building blocks, that is, "co-crystals", e.g. semiconductors and metals or magnetic and fluorescent constituents. For NCL- or NCR-only co-crystals, the examples summarised in section 2.2. may hopefully motivate to develop the field further into this direction. The sub-field of mixed NCL/NCR-co-crystals is currently merely existent, although such combinations may be particularly useful in mitigating some of the challenges of the pure building blocks detailed in section 5.1.

The investigation of emergent mechanical properties is not only relevant for the design of novel stress sensors,[247, 248] but will also help managing mechanical stress in SCs introduced during self-assembly, which is a major source of defects in the SCs. NCL assembled frameworks with intercluster linkers have enhanced properties and stability. NCLs form assemblies with different functional materials such as fullerene, crown ethers, MOFs.[249-251] Their co-crystallization can create materials with novel properties that can be integrated into optoelectronic devices.[42]

Anisotropic charge transport should be a general feature of SCs in the weak coupling regime (as detailed in section 4.2), and further studies in this direction are likely to advance our fundamental understanding of charge carrier transport in such materials, which are also referred to as "granular metals".[252] In addition, anisotropic charge transport can be technologically relevant, e.g. by exploiting direction-dependent electron vs. hole transport for exciton fission and reduced electron-hole recombination, which is important for photovoltaic applications.[253]

Perhaps the largest technological potential for SCs lies in their emergent optical properties, in particular for electro-optical modulation.[254] For instance, the excellent structural integrity and atomically precise composition in combination with crystal structures of low symmetry render SCs of NCLs potentially birefringent. As such, they may find application for refractive index modulations by exploiting the Pockels- or Kerr-effect, e.g. in waveplates. Since these applications are typically limited to photons with sub-bandgap energies to enable near-zero absorption by the birefringent material, the inherent photo-instability of NCLs is of little concern for such applications.

NCRs tolerate much higher resonant photoexcitation and, consequently, SCs composed of these building blocks may find application as electro-optical modulators by exploiting the electrical or optical quantum confined Stark effect under resonant excitation. Previous research efforts in this direction have typically used disordered ensembles of NCRs, which were reasonably narrow in size-distribution, but still lacked the uniformity of highly ordered SCs. We foresee great potential in applying NCR SCs toward this end, since the "self-cleaning" effect in SCs (e.g. further narrowing of the size-distribution)[245] may overcome the current challenges for NCRs as modulators due to spectral broadening, and simultaneously unfold the full advantage of zero-dimensional quantum objects over the currently applied two-dimensional quantum wells.[239]

## 6 Conclusions

We have shown an ample number of examples for well-documented analogies between atoms in a crystal and NCRs/NCLs in an SC. This feature article would not be complete without pointing out some of the limitations of this analogy: Our understanding of the properties of atomic crystals is largely based on the linear combination of atomic orbitals (LCAO), which requires, in essence, strong coupling. As detailed in section 2.3, strong coupling in highly ordered SCs has so far not been achieved, such that an "LCAO-like" description of SCs is not realistic. It is, thus, not surprising that all the "quasi-atomic" properties of NCRs/NCLs listed in section 4 are of substantially different origin than their analogues in atomic crystals. For instance, the transport anisotropy in graphite is based on strong in-plane coupling between $sp^2$-orbitals and weak out-of-plane coupling between $p_z$-orbitals, leading to a 2.3-times shorter C-C bond length in the in-plane direction and the delocalisation of π-electrons. In diamond, the C-C bond length is even shorter than in the in-plane direction of graphite and coupling is strong, but the absence of any free electrons due to the filled valence- and empty conduction band renders diamond an insulator. In SCs of PbS NCRs however, free electrons are always present due to defect-induced doping of the NCRs. Coupling of these electrons is weak in all directions, which enforces the effect of the interparticle distance, and anisotropy emerges from the different interparticle spacings dictated by the unit cell of the SC (section 4.2).

Surface reconstruction in atomic crystals is driven by the undercoordination of the atoms on the surface and empty atomic orbitals or orbitals with unpaired electrons. The geometric rearrangements that occur to minimize the energy of these atoms typically affect only the first few layers near the surface. In contrast, the structural deteriorations in SCs shown in section 4.3 radiate deep into the interior of the SC over length scales of 100 layers and more. Orbital-like interactions do not play a dominant role in SCs (since coupling is weak), such that the undercoordination of nanocrystals at the surface is not as decisive as in atomic crystals. Instead, residual lattice stress, introduced e.g., due to size inhomogeneities of the NCRs within the SC dominates, which is not a factor in atomic crystals grown under thermodynamic conditions.

An adequate reply to Goubet's and Pileni's question quoted in the beginning of this article could therefore be: Analogies between atomic crystals and supercrystals are real, but their emergent properties manifest for fundamentally different reasons.

## Bibliographic Sketches

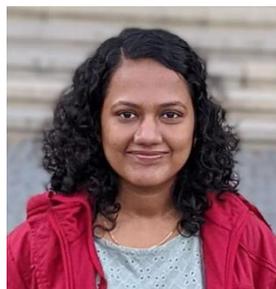

*Korath Shivan Sugi received her Ph.D. from the Indian Institute of Technology Madras in 2021 under the supervision of Prof. Thalappil Pradeep. She joined the University of Tübingen in 2021 for postdoctoral research with Prof. Marcus Scheele. Her research interests focus on assembling atomically precise nanoclusters and their emerging properties.*

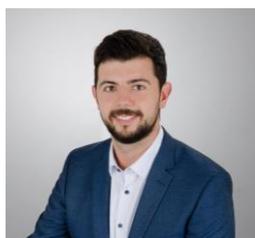

*Andre Maier obtained his PhD at the University of Tübingen in 2021 with Prof. Marcus Scheele. He focussed on the investigation of structure-transport correlations in self-assembled materials of nanocrystals and nanoclusters. Since then, he continued as a postdoc to work on the development of novel materials for ultrafast photodetectors and their characterization methods.*

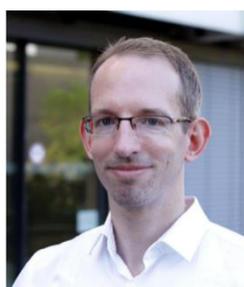

*Marcus Scheele received his PhD from the University of Hamburg in 2011. After a postdoctoral visit at UC Berkeley, he led an Emmy Noether group at the University of Tuebingen and was appointed as a Heisenberg Professor for Physical Chemistry in 2020. His group investigates light-matter interactions in nanomaterials and their application for optical communication or in light emitting diodes.*

## Author Contributions
The manuscript was written by all authors.

## Conflicts of interest
There are no conflicts to declare.

## Acknowledgements
This work was supported by the DFG (grant SCHE1905/9-1), the Carl Zeiss Stiftung (Forschungsstrukturkonzept "Interdisziplinäres nanoBCP-Lab") as well as the European Research Council (ERC) under the European Union's Horizon 2020 research and innovation program (grant agreement No 802822).

## Notes and references